\documentclass[twoside,a4paper]{article}
\usepackage[english]{babel}
\usepackage[letterpaper,top=3cm,bottom=3cm,left=3cm,right=3cm,marginparwidth=1.75cm]{geometry}

\usepackage{amsmath,graphicx,amssymb,fancyhdr,amsthm,tensor}

\theoremstyle{definition}

\theoremstyle{remark}

\def\beq{\begin{eqnarray}}
\def\eeq{\end{eqnarray}}
\def\bsp{\begin{split}}
\def\esp{\end{split}}

\usepackage{tensor}

\usepackage{changepage}

\usepackage{accents}

\usepackage[colorlinks=true, allcolors=blue]{hyperref}

\usepackage{subcaption}

\usepackage{flexisym}
\usepackage{breqn}

\newcommand{\tm}{\tilde{m}}
\newcommand{\tv}{\tilde{\varphi}}

\title{Black Hole Persistence in Scalar Tensor Theories}
\author{B. Yildirim and A. A. Coley}

\begin{document}

\begin{center}
{\Large \bf Black Hole Persistence in Scalar Tensor Theories\\ }
\vskip 0.7cm
{B. Yildirim and A. A. Coley}
\vskip 0.2cm
{\it Department of Mathematics and Statistics\\ 
Dalhousie University\\
Halifax, Canada}\\
\vskip 1cm

\begin{quote}
{\bf Abstract.}~{\small We construct a perturbative scalar-tensor solution describing a central inhomogeneity embedded in an evolving cosmological background, with the aim of studying black hole persistence through a nonsingular bounce. Scalar-tensor gravity provides a natural framework for realizing bouncing cosmologies, while the inclusion of a localized inhomogeneity makes the field equations substantially more difficult to solve. We therefore adopt a perturbative scheme, with perturbative parameter $\epsilon$, in which the leading-order equations are solved by a spatially flat bouncing FLRW spacetime sourced by a radiation perfect fluid. At next order, a central inhomogeneity is introduced through a generalized McVittie geometry, with the perturbations encoded in the corresponding first-order metric and scalar-field functions. We first allow an anisotropic fluid with radial and tangential pressures, whose diagonal components solve the diagonal field equations. The field equations are solved as a series expansion up to $\mathcal{O}(\eta^4)$ near the bounce at $\eta=0$. The resulting perfect fluid solution contains three arbitrary functions which are constrained by requiring the spacetime to asymptote to FLRW as $r\to\infty$. With suitable initial conditions preserving the parabolic structure of the bounce, the integration constant $d_0$ emerges as the true perturbative parameter: all perturbations vanish as $d_0\to0$. Finally, we find a small evolving horizon, $r_h\sim d_0$, which we interpret as the horizon of the central inhomogeneity. Its persistence through the bounce supports the interpretation of a black hole surviving the cosmological transition, and its evolution is not symmetric about $\eta=0$.

}
\end{quote}
\end{center}
\vskip 1.0cm

\noindent Keywords: Friedmann-Lemaıtre-Robertson-Walker (FLRW). General Relativity (GR). Scalar Tensor. Black Hole Persistence.

\section*{Author contributions}

All authors contributed to the research. Balkar Yildirim wrote the first draft of the manuscript. All authors reviewed and provided edits to the manuscript.

\newpage

\section{Introduction}

 An alternative cosmology to the standard $\Lambda$CDM cosmology based on the Friedmann-Lemaître-Robertson-Walker (FLRW) model, which is currently of interest, is a non-singular bouncing cosmology \cite{novello}, in which a previously contracting era connects with a presently expanding era by means of some minimal scale factor (and hence a vanishing Hubble rate). These models were originally proposed in \cite{Tolman31,Lemaitre33,Lemaitre:1933gd}. The bounce, which connects these two eras, can be generated by classical effects such as with a cosmological constant \cite{lemaitre}, semi-quantum gravitational effects (e.g., associated with string theory \cite{string}\cite{Cai:2009in}, the ``pre-big-bang'' scenario \cite{BHScorr}, loop quantum gravity \cite{sing} \cite{WilsonEwing:2012pu}), and quantum cosmological effects \cite{ashtekar} \cite{hertog}.  Bouncing cosmologies were reviewed in \cite{brand,novello,Battefeld_2015,YCL}). 

The mechanics of producing a bounce is to violate one of the energy conditions \cite{Battefeld_2015}. There are essentially two ways to effectively achieve this: the first requires breaking of the null energy conditions due to matter fields with negative kinetic energy \cite{Martin:2003sf}. The second allows for a classically singular bounce via an alternative theory of gravity (i.e.,  a modification to Einstein's field equations of General Relativity (GR)), where the geometric additions to GR effectively break the null energy condition while the matter still obeys it.

There is the interesting possibility~\cite{cc} that black holes can persist in a universe which re-collapses to a big crunch and subsequently bounces into a new expansion phase \cite{cc}; these types of black holes are called ``pre-big-bang black holes". This allows the possibility that some of the black holes in the present era originated in the previous era and survived the bounce. These would be distinct from the ``primordial black holes'' which formed just after the big bang. If primordial black holes or pre-big-bang black holes were big enough, they would still exist today and they could be the solution to many cosmological puzzles, from dark matter to supermassive black holes \cite{cc}. The possibility that dark matter could consist of primordial black holes has become topical \cite{carr-silk}. In a variety of mass ranges, pre-big-bang black holes could also contribute to dark matter, provide seeds for galaxies, generate entropy and even drive the bounce itself.

Rovelli and Vidotto~\cite{rovelli} have studied the possibility that dark matter is composed of the remnants of pre-big-bang black holes. This then leads to the question as to whether the same black holes could make up the dark matter in successive cosmic cycles, with the fraction of dark matter even progressively increasing. Recently \cite{cainew} it was suggested that it might eventually be possible to distinguish pre-big-bang black holes from black holes formed in our current universe since we know that they already existed very early, and it is not apparent how they could grow so big so fast unless they were seeded. 

In \cite{ccc} a class of four-dimensional dynamical exact solutions describing a regular lattice of black holes in a cosmological background dominated by a scalar field at the bounce was obtained. These solutions show that exact configurations can exist in which several black holes survive through a bounce. Since the bounce is not necessarily expected to take place at the Planck scale, a classical treatment remains well justified.

More recently, Corman et al.\cite{corman} numerically analyzed the evolution of a black hole through a non-singular cosmological bounce in a model sourced by both a ghost scalar field, used as an effective mechanism for violating the energy conditions, and an ordinary scalar field. By solving the nonlinear Einstein field equations, they followed black holes of different sizes across the bounce. In particular, asymptotically cosmological initial data were chosen to generate a contracting phase, followed by a bounce and then by cosmological expansion. It was shown in \cite{quintin} that black holes formed, or already present, during the contracting phase can persist through the bounce and may therefore leave observational signatures in the post-bounce universe. In addition, P{\'e}rez and Romero \cite{perez} studied the behavior of a black hole during a bouncing phase. Although their computational approach differs, both analyses found that the black hole can survive the bounce. It was also shown that the horizon in the model of \cite{perez} evolves with cosmic time, since it is coupled to the background cosmological dynamics.

\subsection{Generalized McVittie solution}

The McVittie spacetime \cite{Mcvittieoriginal} is an exact solution of the Einstein field equations sourced by a perfect fluid, describing a spherically symmetric Schwarzschild black hole embedded in a spatially flat FLRW cosmological background. The global structure of the McVittie spacetime has been studied in \cite{A107,A110}. The original choice of mass function in the McVittie metric \cite{Mcvittieoriginal} is known as the \emph{non-accretion condition}, and follows from solving the off-diagonal field equation $G \indices{^0 _1} = 0$. More recently, Kaloper et al.~\cite{Kaloper_2010} revisited the role of a positive cosmological constant in the spatially flat McVittie solution.

As discussed above, the question of whether a black hole can survive in a universe undergoing a cosmological bounce was investigated based on a generalization of the spatially flat McVittie spacetime \cite{Faraoni_2007} and first analyzed in \cite{perez}. That analysis is limited by the fact that the governing equations are under-determined. As a result, the metric evolution is imposed in an ad hoc way, after which the corresponding imperfect-fluid matter content and its evolution are derived. In addition, a scale-factor ansatz motivated by quantum theory \cite{Peter_2007,Pinto_Neto_2013} was assumed from the outset in order to generate a classical bounce.

More specifically, Perez and Romero \cite{perez} generalized the standard flat McVittie metric by relaxing the non-accretion condition, thereby allowing the central black hole to interact with a dynamical cosmological background. In order for the solution not to reduce to the standard McVittie metric after imposing the field equations, this generalization requires a non-vanishing off-diagonal component of the energy-momentum tensor. The bounce is introduced through ad hoc quantum corrections to the classical cosmological scale factor $a(t)$, taking the form
$a(t) = a_b [ 1 + \left( {t}/{t_b} \right)^2]^{\frac{1}{3}}$.
Perez and Romero specify the metric functions $a(t)$ and $m(t)$ \emph{a priori}, and then obtain the physical quantities by defining them through the resulting field equations. Consequently, the model is not an exact solution in the usual sense. Initially, the choice $m(t) = m_0$ was adopted \cite{perez}, while in later work the form $m(t)=\frac{m_0}{a(t)}$ was assumed \cite{P_rez_2021}.

\subsection{Recent Work}

In our recent paper \cite{YCL}, we relaxed the assumptions made in \cite{perez} and derived an exact FLRW bouncing solution with positive spatial curvature and a positive cosmological constant. No ansatz was imposed for the scale factor; instead, its bouncing behaviour was obtained directly from the field equations of New General Relativity (NGR) \cite{Bahamonde_2023}, without requiring violations of the matter energy conditions.

NGR is a teleparallel extension of GR that replaces curvature with torsion, and it can also admit bouncing solutions through a mechanism closely related to the one appearing in GR for a closed FLRW model with a positive cosmological constant \cite{YCL}. Starting from a generalized McVittie metric, we used a perturbative and local approach was used to describe a bouncing cosmological background containing a central inhomogeneity within a one-parameter NGR theory. At leading order, the effect of NGR is to modify the GR FLRW solution by renormalizing the positive spatial-curvature term in the Friedman equation. Hence, the leading-order behaviour of the perturbed solution is governed by the NGR FLRW background, while the central inhomogeneity enters only at higher perturbative orders.

No additional assumptions were initially imposed on the spacetime, such as a prescribed form for the scale factor. Instead, the local evolution was derived from the field equations using a perturbative scheme valid ``near the bounce''. Thus, the resulting investigation was not intended as a complete global analysis in space or time, but rather as an effective description valid near the bounce in time, namely at very early times, and far from the central inhomogeneity in space.

Qualitatively, the perturbation modifies the minimum of the bounce at $t=0$, while the evolution remains symmetric near the bounce. The central inhomogeneity evolves at higher perturbative orders, with its behaviour depending on the arbitrary constants appearing in the perturbative solution.

The local horizon also undergoes a bounce. The perturbation does not change the minimum value of the horizon, but instead modifies the higher-order terms in its evolution. It is worth noting that the leading-order horizon is symmetric about $t=0$, whereas the perturbative correction breaks this symmetry through the term linear in $t$.

As emphasized in \cite{YCL}, the above analysis applies to any bouncing model in a theory of gravity that locally gives rise to an effective positive spatial curvature and an effective cosmological constant. In particular, this approach is well suited for studying bouncing models in scalar-tensor theories of gravity.

\subsection{Scalar-tensor Theories of Gravity}

We can apply the techniques for studying black hole persistence to a more conventional theory like a scalar-tensor theory. Due to the addition of a scalar-field we can consider spatially flat cosmological solutions while still having a bounce. For example, we can consider the solutions in Generalized Brans-Dicke (GBD) theory given in \cite{barrow_nonsingular_1993} \cite{barrow_scalar-tensor_1993}, where GBD is a generalized version of Brans-Dicke (BD) in which the BD parameter $\omega$ is a function of the scalar field. In particular, we will be interested in the spatially flat solution with a radiation fluid. As before, we will begin with choosing an FLRW bouncing solution, then adding the perturbations via the generalized McVittie metric. A generic scalar-tensor theory is described by the following action \cite{cosmologyinscalar}

\begin{equation}
    S = \int d^4x \sqrt{-g} \left[ F(\phi) R - Z(\phi) \nabla_a \phi \nabla^a \phi - V(\phi) \right]
\end{equation}

\noindent  We can transform this action into a Brans-Dicke like action by defining $\Phi = F(\phi)$ and $\omega(\Phi) = \frac{F(\phi) Z(\phi)}{F'(\phi)^2}$ then the action becomes

\begin{equation}
    S = \frac{1}{16 \pi} \int d^4x \sqrt{-g} \left( \Phi R - \frac{\omega(\Phi)}{\Phi} \nabla^\mu \Phi \nabla_\mu \Phi - V(\Phi) \right) + S^{(m)}
\end{equation}

\noindent where $\Phi$ is some dynamical scalar field determined by the field equations. Variation with respect to the metric and then the scalar field gives the following field equations

\begin{eqnarray}
    G_{\mu \nu} = \frac{8 \pi}{\Phi} T_{\mu \nu} + \frac{\omega(\Phi)}{\Phi^2} \bigg( \nabla_\mu \Phi \nabla_\nu \Phi - \frac{1}{2} g_{\mu \nu} \nabla^\sigma \Phi \nabla_\sigma \Phi \bigg) + \frac{1}{\Phi} \bigg( \nabla_\mu \nabla_\nu \Phi - g_{\mu \nu} \Box \Phi \bigg) - \frac{1}{2 \Phi} g_{\mu \nu } V(\Phi) \label{fe1}
\end{eqnarray}

\begin{equation}
    (2 + 3 \omega(\Phi)) \Box \Phi = 8 \pi T - \omega'(\Phi) \nabla^\sigma \Phi \nabla_\sigma \Phi - 2 V(\Phi) + \Phi V'(\Phi) \label{fe2}
\end{equation}

\section{Analysis}

\noindent At leading order the dynamics will be governed by a FLRW bouncing solution. We will use the spatially flat FLRW metric

\begin{equation}
    ds^2 = - dt^2 + a(t)^2 (dr^2 + r^2 d\Omega^2)
\end{equation}

\noindent with a perfect radiation fluid

\begin{equation}
    T \indices{^\mu _\nu} = \frac{1}{8 \pi} \text{diag} \left( -\frac{3 \Gamma}{a(t)^4},\frac{ \Gamma}{a(t)^4},\frac{ \Gamma}{a(t)^4},\frac{ \Gamma}{a(t)^4}  \right)
\end{equation}

\noindent where $\Gamma$ is a constant. The scalar field will be chosen to be spatially homogeneous to obey the FLRW symmetry, $\Phi =\varphi(t)$. We then obtain the two independent field equations from \eqref{fe1}, \eqref{fe2}

\begin{dgroup}
\begin{dmath}
    \frac{\dot{a}^2}{a^2} = \frac{\Gamma}{\varphi a^4} - \frac{\dot{\varphi}}{\varphi} \frac{\dot{a}}{a} + \frac{\omega(\varphi)}{6} \frac{\dot{\varphi}^2}{\varphi^2} + \frac{V(\varphi)}{\varphi}
\end{dmath}

\begin{dmath}
    \ddot{\varphi} + \frac{3 \dot{a}}{a} \dot{\varphi} = -\frac{\dot{\varphi}^2 \omega'(\varphi)}{3+2\omega(\varphi)} +\frac{2V(\varphi) + \varphi V'(\varphi)}{3+2\omega(\varphi)}
\end{dmath}
\end{dgroup}

\noindent To solve these equations we will use the solutions provided in \cite{barrow_nonsingular_1993} which uses the conformal time coordinate $\eta$, which is related to the cosmological time $t$ by $a d\eta = dt$. In particular, we will be concerned with the radiation fluid solution where $\Gamma > 0$, which is given by

\begin{dgroup}[label={flrwscalescalar}]
\begin{dmath}
    a(\eta) = \sqrt{\frac{1}{\varphi_c} \Gamma \left( -\frac{A^2}{4 \Gamma^2} + (\eta + \eta_0)^2 \right) \left( 1 + f^{-\lambda} \right)} \label{radasc}
\end{dmath}

\begin{dmath}
    \varphi(\eta) = \varphi_c f^\lambda (1+f^\lambda)^{-1} \label{radphi}
\end{dmath}
\end{dgroup}

\noindent where $A$,$\eta_0$,$\varphi_c$ are constants and we have
\begin{eqnarray}
    \omega(\varphi) &=& \frac{\beta}{\left( 1- \frac{\varphi}{\varphi_c} \right)^2} - \frac{3}{2}, \quad V(\varphi) = 0 \\
    f &\equiv& \frac{\eta + \eta_0 - \frac{A}{2 \Gamma}}{\eta + \eta_0 + \frac{A}{2 \Gamma}}, \quad \lambda \equiv \sqrt{\frac{3}{2 \beta}}
\end{eqnarray}

\noindent This solution exhibits a bounce for $\lambda > 1$ \cite{barrow_nonsingular_1993}. 

\subsection{Minimum of Bounce} \label{sec2.1}

We want to find the location of the minimum of the scale factor and choose $\eta_0$ such that the minimum occurs at $\eta=0$. We can do this by asserting $\dot{a}(0) = 0$ and solving for $\eta_0$. This requires us to solve the following equation for $\eta_0$

\begin{equation}
    2 \Gamma  \eta_0 \left( 1 +s^{\lambda }\right)-A \lambda = 0 \label{radmincond}
\end{equation}

\noindent where

\begin{equation}
    s^\lambda \equiv \left(1-\frac{2 A}{A+2 \Gamma \eta_0}\right)^\lambda
\end{equation}

\noindent This equation cannot be solved for $\eta_0$ exactly for arbitrary $\lambda > 1$. We shall carry the constant $\eta_0$ forward, but it will need to be estimated for any application. We can solve \eqref{radmincond} implicitly by asserting the condition

\begin{dmath}
    s^\lambda = \frac{A\lambda}{2\Gamma \eta_0}-1 \label{implicitcond}
\end{dmath}

\noindent which satisfied \eqref{radmincond}. This allows us to expand the scale factor and scalar field \eqref{flrwscalescalar} about $\eta=0$:

\begin{dmath}
    a(\eta) = \left[\frac{1}{2 \Gamma} \sqrt{\frac{A \Gamma \lambda \left( A^2-4\Gamma^2\eta_0^2\right)}{\varphi_c (2 \eta_0 \Gamma- A \lambda)}} \right] + \left[ \frac{\Gamma^2 \left( 4A\eta_0\Gamma \lambda - A^2 - 4 \eta_0^2 \Gamma^2 \right)}{\left( A^2 - 4\eta_0^2 \Gamma^2 \right)^2} \sqrt{\frac{A \Gamma \lambda \left( A^2-4\Gamma^2\eta_0^2\right)}{\varphi_c (2 \eta_0 \Gamma- A \lambda)}} \right] \eta^2 + \mathcal{O}(\eta^3)
\end{dmath}

\begin{dmath}
    \varphi(\eta) = \left[ \varphi_c \left( 1 - \frac{2\eta_0 \Gamma}{A \lambda} \right) \right] + \left[ \frac{8 \eta_0 \varphi_c \Gamma^2 \left( 2\eta_0 \Gamma - A \lambda \right)}{A \lambda \left( A^2 - 4 \eta_0^2 \Gamma^2 \right)} \right] \eta + \left[- \frac{16 \eta_0 \varphi_c \Gamma^3 \left( 2\eta_0 \Gamma - A \lambda \right)}{A \lambda \left( A^2 - 4 \eta_0^2 \Gamma^2 \right)} \right] \eta^2 + \mathcal{O}(\eta^3)
\end{dmath}

\noindent Looking at the $\eta^0$ order term of $a(\eta)$ we can impose conditions on the parameters such that the scale factor is real. First, to have a non singular bounce, we require the scale factor to be of form $a \sim a_0 + a_2 \eta^2$ where $a_0,a_2 > 0$. Hence, $A \neq 2 \eta_0 \Gamma$ and $A \neq 2 \eta_0 \Gamma/\lambda$. First, $\Gamma>0$ and assuming $\eta_0 > 0$ its clear that $2 \eta_0 \Gamma/\lambda < 2 \eta_0 \Gamma$ for $\lambda > 1$. Therefore, we have three options

\begin{enumerate}
    \item $A < \frac{2 \eta_0 \Gamma}{\lambda} < 2 \eta_0 \Gamma \implies \frac{A}{\varphi_c} <0 $
    \item $ \frac{2 \eta_0 \Gamma}{\lambda} < A < 2 \eta_0 \Gamma \implies \frac{A}{\varphi_c} > 0$
    \item $ \frac{2 \eta_0 \Gamma}{\lambda} <  2 \eta_0 \Gamma < A \implies \frac{A}{\varphi_c} <0$
\end{enumerate}

\noindent And there is a similar set of conditions for $\eta_0 < 0$. Note that $\eta_0 \neq 0$ as this would contradict the implicit condition for the bounce to occur at $\eta=0$ \eqref{implicitcond}.  

To have a bounce we require $a_2 > 0$, which implies

\begin{dmath}
     4A\eta_0\Gamma \lambda - A^2 - 4 \eta_0^2 \Gamma^2 > 0
\end{dmath}

\noindent We can rewrite this inequality as 

\begin{equation}
    \frac{\lambda-\sqrt{\lambda^2-1}}{2} < \frac{\eta_0 \Gamma}{A} < \frac{\lambda+\sqrt{\lambda^2-1}}{2}
\end{equation}

\noindent Then there exists values of $\frac{\eta_0 \Gamma}{A}>0$ that satisfy this equality only when $\lambda > 1$, which is indeed the condition for a bounce to occur as expected.

Finally, even though we do not need to exactly find the value of $\eta_0$ such that $\dot{a}(0)=0$ we can still attempt to find an approximate value for general $\lambda$. The expression that is preventing us from solving exactly is $s^\lambda$ and so we must find a way to approximate this expression. There is no obvious method as the constant parameters $\Gamma$, $A$ and $\eta_0$ are arbitrary. We can get a hint by solving \eqref{radmincond} for $\eta_0$ exactly for different values of $\lambda$. We find 

\begin{equation}
    \eta_0^{\lambda=2} = \frac{A}{\Gamma} (0.92...), \quad \eta_0^{\lambda=3} = \frac{A}{\Gamma} (1.36...), \quad \eta_0^{\lambda=4} = \frac{A}{\Gamma} (1.81...)
\end{equation}

\noindent We can see that $\frac{\Gamma}{A}\eta_0 \approx 1$. We can then define the parameter

\begin{equation}
    X \equiv 1 - \frac{\Gamma}{A}\eta_0 \approx 0 
\end{equation}

\noindent Note that as $\lambda$ gets larger so does $\eta_0$, hence this approximation ($X \approx 0$) is only valid strictly for $\lambda \sim 1$. Regardless, using this parameter $X$, we have that

\begin{equation}
    s^\lambda = \left( 1-\frac{2}{3-2X} \right)^\lambda
\end{equation}

\noindent and for $X \approx 0$ using the binomial approximation we have

\begin{equation}
    s^\lambda \approx 3^{-(1+\lambda)} \left( 3- 4 \lambda X \right)
\end{equation}

\noindent Substituting the approximate form of $s^\lambda$ back into \eqref{radmincond} and expressing in terms of $X$, we now need to solve

\begin{equation}
    1 +3^{-(1+\lambda)} \left( 3- 4 \lambda X \right)- \frac{\lambda}{2(1-X)} = 0 \label{radmincondapprox}
\end{equation}

\noindent which is a quadratic equation for $X$. Solving for $X$ we have

\begin{equation}
    X = \frac{1}{8 \lambda} \left( 3(1+3^\lambda)+4\lambda \pm \sqrt{9(1+3^\lambda)^2+ 8\lambda (2\lambda-3 + 3^{1+\lambda}(\lambda-1))} \right)
\end{equation}

\noindent This then gives

\begin{equation}
    \eta_0^{\lambda=2} \approx \frac{A}{\Gamma} (0.919...), \quad \eta_0^{\lambda=3} \approx \frac{A}{\Gamma} (1.373...), \quad \eta_0^{\lambda=4} \approx \frac{A}{\Gamma} (1.869...)
\end{equation}

In the event that this minimum occurs at $\eta=0$, then in terms of cosmological time $t$ the minimum will occur at $t=0$. We can see this by using the definition of conformal time

\begin{equation}
    t = \int a(\eta) d\eta 
    = \int a_0 + a_2 \eta^2 + ... d\eta 
    = a_0 \eta + a_2 \frac{\eta^3}{3} + t_0 ... 
\end{equation}

\noindent which is zero when $\eta=0$ if we choose the integration constant $t_0=0$.

\subsection{Setup of Perturbative Scheme}

A bouncing cosmological solution is achieved by beginning with a FLRW solution. Utilizing the McVittie metric, which can be interpreted as a central inhomogeneity within a cosmological background, together with its FLRW limit when the mass function goes to zero, we will investigate the following (generalized) McVittie metric:

\begin{equation}
    ds^2 = - \frac{\left( 1 - \frac{M(\eta,r)}{2r} \right)^2}{\left( 1 + \frac{M(\eta,r)}{2r} \right)^2} a(\eta)^2 d \eta^2 +A(\eta,r)^2 \left( 1 + \frac{M(\eta,r)}{2r} \right)^4 \left( dr^2 + r^2 d\Omega^2 \right) \label{mcvittie}
\end{equation}

\noindent It may seem strange that we have used the conformal time here as well; however, we can at anytime return to the cosmological time and recover the standard form of this metric. The energy-momentum tensor is given by

\begin{equation}
    T \indices{^\mu _\nu} = \text{diag} (-\rho(t,r), p_r(t,r), p_t(t,r), p_t(t,r))
\end{equation}

\noindent Notice that we have generalized the flat McVittie metric by including the metric functions $A$ and $M$ which are assumed to be both functions of $\eta$ and $r$. In general, the energy-momentum tensor models an inhomogeneous imperfect fluid with radial and tangential pressures. However, we will consider a perfect fluid solution later on. This metric is spherically symmetric and has the FLRW limit when $M(\eta,r)/r \to 0$ and $A(\eta,r) \to A(\eta)$. To this end, we will choose the following perturbative ansatz:

\begin{eqnarray}
    A(\eta,r) &=& a(\eta) + \epsilon \tilde{a}(\eta,r) \label{pertansatzA}\\
    M(\eta,r) &=& 0 + \epsilon \tilde{m}(\eta,r) \\
    \rho(\eta,r) &=& \tfrac{3\Gamma}{8 \pi a(\eta)^4} + \epsilon \tilde{\rho}(\eta,r) \\
    p_r(\eta,r) &=& \tfrac{\Gamma}{8 \pi a(\eta)^4} + \epsilon \tilde{p}_r(\eta,r) \\
    p_t(\eta,r) &=& \tfrac{\Gamma}{8 \pi a(\eta)^4} + \epsilon \tilde{p}_t(\eta,r) \\ 
    \Phi(\eta,r) &=& \varphi(\eta) + \epsilon \tilde{\varphi}(\eta,r) \label{pertansatzPhi}
\end{eqnarray}

 \noindent Here $\epsilon$ is a dummy perturbative parameter --- it only keeps track of the order of the terms and will eventually be set to $\epsilon=1$. Notice that when $\epsilon=0$, the metric \eqref{mcvittie} becomes FLRW. We will then expand the field equations of \eqref{mcvittie} in terms of $\epsilon$ up to $\mathcal{O}(\epsilon)$ using the above perturbed ansatz. 
 
 The FLRW solutions discussed previously will determine the form of the leading order functions and the perturbations at order $\epsilon^1$ will be deduced from the field equations. In addition, the FLRW solution has no potential; we will choose the non-perturbative potential to be of form

\begin{equation}
    V(\Phi) = 0 + \epsilon \tilde{V}(\Phi) \label{pertansantzV}
\end{equation}

\noindent Since $\Phi$ itself is expanded in terms of $\epsilon$, we obtain

\begin{equation}
    V(\Phi) = \epsilon  \tilde{V} (\varphi(\eta)) + \mathcal{O}(\epsilon^2)
\end{equation}

\noindent Hence, we introduce some spatially homogeneous potential at $\mathcal{O}(\epsilon)$ while maintaining a vanishing potential at the leading order $\epsilon^0$. And finally, we will choose the function $\omega(\Phi)$ according to the FLRW solution discussed above. The $\Phi$ in question will itself be perturbed similar to the potential. I.e., $\omega(\Phi) = \omega(\varphi) + \epsilon \omega'(\varphi) \varphi + \mathcal{O}(\epsilon^2)$.

\begin{equation}
    \omega(\Phi) = \frac{\beta}{\left( 1- \frac{\Phi}{\varphi_c} \right)^2} - \frac{3}{2}
\end{equation}

At $\mathcal{O}(\epsilon)$, using the generalized McVittie metric we will obtain six equations, including the condition that we have a perfect fluid at $\mathcal{O}(\epsilon)$, but we have seven unknown functions. However, $\tilde{V}$ will not be treated as dynamical (as in, not the solution of a differential equation), but will be kept for generality or potentially chosen in such a way that simplifies the field equations at the perturbative level. Therefore, we will solve for the tilde functions with $\mathcal{O}(\epsilon)$ field equations.

For convenience, all terms of the field equations will be placed on the left hand side. That is, 

\begin{equation}
    E_{\mu \nu} \equiv G_{\mu \nu}- \frac{1}{8 \pi \Phi} T_{\mu \nu} + (\Phi \\\ \text{terms}) 
\end{equation}

\noindent then $E_{\mu \nu}=0$ expresses the field equations (likewise with the $\Phi$ field equation). Expanding in terms of $\epsilon$ we have

\begin{equation}
    E_{\mu \nu}=E^{(0)}_{\mu \nu} + \epsilon E^{(1)}_{\mu \nu} + \mathcal{O}(\epsilon^2) = 0
\end{equation}

\noindent Then we will solve $E^{(i)}_{\mu \nu}=0$ for each order up to $\mathcal{O}(\epsilon)$ ($i=0,1$) omitting all $E^{(i)}_{\mu \nu}$ for $i \geq 2$ as negligible. Since we have already chosen the $\mathcal{O}(\epsilon^0)$ part of the ansatz \eqref{pertansatzA}-\eqref{pertansantzV} such that $E^{(0)}_{\mu \nu}=0$ are already satisfied with the FLRW bounce solutions discussed in the previous section, all that is left is to solve for the corrections to the FLRW solution by solving $E^{(1)}_{\mu \nu}=0$ for the functions in the perturbative ansatz with a tilde.

\subsection{Field Equations}

First, we will list the non-zero components of $E^{(1)}_{\mu \nu}=0$. We will use the dot to mean the derivative with respect to $\eta$ and the prime to mean the derivative with respect to $r$.

\begin{multline}
    E^{(1)}_{(00)} =  a^5 r \varphi ^2 (16 \pi  \tilde{\rho}+\tilde{V}(\varphi ))+a^3 \left(2 \varphi  \left(\varphi  \left(r \tilde{\varphi}''+2 \varphi  \tilde{m}''+2 \tilde{\varphi}'\right)+\dot{\varphi } \left(\omega (\varphi ) \left(r \dot{\tilde{\varphi}}+\tilde{m} \dot{\varphi }\right)-3 \dot{\tilde{m}} \varphi \right)\right) \right. \\ \left. +r \tilde{\varphi} \dot{\varphi }^2 \left(\varphi  \omega '(\varphi )-2 \omega (\varphi )\right)\right)+2 a^2 \varphi  \left(3 \dot{a} r \tilde{\varphi} \dot{\varphi }-\varphi  \left(-2 \varphi  \left(-3 \dot{a} \dot{\tilde{m}}+r \tilde{a}''+2 \tilde{a}'\right)+3 r \left(\dot{a} \dot{\tilde{\varphi}}+\dot{\tilde{a}} \dot{\varphi }\right) \right. \right. \\ \left. \left. +6 \dot{a} \tilde{m} \dot{\varphi }\right)\right) +6 \dot{a} a \varphi ^2 \left(r \tilde{a} \dot{\varphi }-2 \varphi  \left(\dot{a} \tilde{m}+r \dot{\tilde{a}}\right)\right)+12 \dot{a}^2 r \tilde{a} \varphi ^3 = 0  \label{E(00)}
\end{multline}

\begin{multline}
    E^{(1)}_{(11)} =  a^4 r \varphi ^2 (\tilde{V}(\varphi )-16 \pi  \tilde{p}_r)+4\varphi ^2\left(\dot{a}^2 \tilde{m} \varphi +r\tilde{a}\left(\dot{a} \dot{\varphi }+\varphi \ddot{a}\right)\right)-2a\varphi \left(\dot{a} (-r) \tilde{\varphi} \dot{\varphi }+\varphi \left(2\tilde{m}\left(\dot{a} \dot{\varphi } \right. \right. \right. \\ \left. \left. \left. +2\varphi \ddot{a}\right) +2 r \dot{\tilde{a}} \dot{\varphi }+\dot{a} r \dot{\tilde{\varphi}}+\varphi \left(-2 \tilde{a}'+6 \dot{a} \dot{\tilde{m}}+2r\ddot{\tilde{a}}\right)\right)\right)+a^2\left(r\tilde{\varphi}\left(\dot{\varphi }^2 \left(2 \omega (\varphi )-\varphi  \omega '(\varphi )\right)+2\varphi \ddot{\varphi }\right) \right. \\ \left.  -2\varphi \left(\tilde{m}\left(\dot{\varphi }^2 \omega (\varphi )+2\varphi \ddot{\varphi }\right)+r \dot{\tilde{\varphi}} \dot{\varphi } \omega (\varphi )+2\varphi ^2\ddot{\tilde{m}}+\varphi \left(-2 \tilde{\varphi}'+3 \dot{\tilde{m}} \dot{\varphi }+r\ddot{\tilde{\varphi}}\right)\right)\right) = 0 \label{E(11)}
\end{multline}

\begin{multline}
    E^{(1)}_{(22)} =  a^4 r \varphi ^2 (\tilde{V}(\varphi )-16 \pi  \tilde{p}_t)+4\varphi ^2\left(\dot{a}^2 \tilde{m} \varphi +r\tilde{a}\left(\dot{a} \dot{\varphi }+\varphi \ddot{a}\right)\right)+2a\varphi \left(\dot{a} r \tilde{\varphi} \dot{\varphi }-\varphi \left(2\tilde{m}\left(\dot{a} \dot{\varphi }+2\varphi \ddot{a}\right) \right. \right. \\ \left. \left. +r \left(\dot{a} \dot{\tilde{\varphi}}+2 \dot{\tilde{a}} \dot{\varphi }\right)-\varphi \left(\tilde{a}'+r \tilde{a}''-6 \dot{a} \dot{\tilde{m}}-2r\ddot{\tilde{a}}\right)\right)\right)+a^2\left(r\tilde{\varphi}\left(\dot{\varphi }^2 \left(2 \omega (\varphi )-\varphi  \omega '(\varphi )\right)+2\varphi \ddot{\varphi }\right) \right. \\ \left. -2\varphi \left(\tilde{m}\left(\dot{\varphi }^2 \omega (\varphi )+2\varphi \ddot{\varphi }\right)+r \dot{\tilde{\varphi}} \dot{\varphi } \omega (\varphi )+2\varphi ^2\ddot{\tilde{m}}-\varphi \left(\tilde{\varphi}'+r \tilde{\varphi}''-3 \dot{\tilde{m}} \dot{\varphi }-r\ddot{\tilde{\varphi}}\right)\right)\right) = 0 \label{E(22)}
\end{multline}

\begin{multline}
    E^{(1)}_{(01)} =  -\frac{2 \varphi ^2}{r} \left(a^2 \dot{\tilde{m}}+r \left(\dot{a} r \tilde{a}'-a \left(a \dot{\tilde{m}}'+r \dot{\tilde{a}}'\right)\right)\right)+a^2 r \dot{\varphi } \tilde{\varphi}' \omega (\varphi ) \\ +\frac{a \varphi}{r}  \left(2 \dot{a} \varphi +a \dot{\varphi }\right) \left(r \tilde{m}'-\tilde{m}\right)+a r \varphi  \left(a \dot{\tilde{\varphi}}'-\dot{a} \tilde{\varphi}'\right) = 0 \label{E(01)}
\end{multline}

\noindent And finally the field equation for $\Phi$ at $\mathcal{O}(\epsilon)$
\begin{multline}
      -8 \pi ( \tilde{p}_r+2 \tilde{p}_t-\tilde{\rho})+2 \tilde{V}(\varphi )+\varphi  \tilde{V}'(\varphi )-\frac{2\tilde{\varphi}\omega '(\varphi )}{a^3}\left(2 \dot{a} \dot{\varphi }+a\ddot{\varphi }\right) \\ -\frac{\dot{\varphi }}{a^2} \bigg(\dot{\varphi } \left(\frac{2 \tilde{m} \omega '(\varphi )}{r}+\tilde{\varphi} \omega ''(\varphi )\right) +2 \dot{\tilde{\varphi}} \omega '(\varphi )\bigg) +\frac{2 \omega (\varphi )+3}{a^4 r}\bigg(3 \dot{a} r \tilde{a} \dot{\varphi }-a\left(2\tilde{m}\left(2 \dot{a} \dot{\varphi }+a\ddot{\varphi }\right)+r \left(2 \dot{a} \dot{\tilde{\varphi}}+3 \dot{\tilde{a}} \dot{\varphi }\right) \right. \\ \left. +a\left(-2 \tilde{\varphi}'-r \tilde{\varphi}''+4 \dot{\tilde{m}} \dot{\varphi }+r\ddot{\tilde{\varphi}}\right)\right)\bigg) = 0 \label{phiFEepsilon}
\end{multline}

\noindent Then we notice that \eqref{E(00)} - \eqref{E(22)} can be algebraically solved for $\tilde{\rho}, \tilde{p}_r$ and $\tilde{p}_t$, respectively, since their derivatives do not appear in the corresponding equations. We have

\begin{dmath}
    \tilde{\rho} = \frac{1}{16 \pi  a^5 r \varphi ^2} \left(-a^5 r \varphi ^2 \tilde{V}(\varphi )+a^3 \left(r \tilde{\varphi} \dot{\varphi }^2 \left(2 \omega (\varphi )-\varphi  \omega '(\varphi )\right)-2 \varphi  \left(\varphi  \left(r \tilde{\varphi}''+2 \varphi  \tilde{m}''+2 \tilde{\varphi}'\right)+\dot{\varphi } \left(\omega (\varphi ) \left(r \dot{\tilde{\varphi}}+\tilde{m} \dot{\varphi }\right)-3 \dot{\tilde{m}} \varphi \right)\right)\right)+2 a^2 \varphi  \left(\varphi  \left(-2 \varphi  \left(-3 \dot{a} \dot{\tilde{m}}+r \tilde{a}''+2 \tilde{a}'\right)+3 r \left(\dot{a} \dot{\tilde{\varphi}}+\dot{\tilde{a}} \dot{\varphi }\right)+6 \dot{a} \tilde{m} \dot{\varphi }\right)-3 \dot{a} r \tilde{\varphi} \dot{\varphi }\right)+6 a \varphi  \left(\dot{a} \varphi  \left(2 \varphi  \left(\dot{a} \tilde{m}+r \dot{\tilde{a}}\right)-r \tilde{a} \dot{\varphi }\right)+\Gamma  r \tilde{\varphi}\right)-12 \dot{a}^2 r \tilde{a} \varphi ^3 \right) \label{gentrho}
\end{dmath}

\begin{dmath}
    \tilde{p}_r = \frac{1}{16 \pi  a^4 r \varphi ^2}\left(r\tilde{\varphi}\left(2 a^2 \dot{\varphi }^2 \omega (\varphi )+\varphi \left(2 \Gamma +2 a \dot{a} \dot{\varphi }+a^2\left(-\dot{\varphi }^2 \omega '(\varphi )+2\ddot{\varphi }\right)\right)\right)-\varphi \left(-a^4 r \varphi  \tilde{V}(\varphi )-4\varphi \left(\dot{a}^2 \tilde{m} \varphi +r\tilde{a}\left(\dot{a} \dot{\varphi }+\varphi \ddot{a}\right)\right)+2a\varphi \left(2\tilde{m}\left(\dot{a} \dot{\varphi }+2\varphi \ddot{a}\right)+2 r \dot{\tilde{a}} \dot{\varphi }+\dot{a} r \dot{\tilde{\varphi}}+\varphi \left(-2 \tilde{a}'+6 \dot{a} \dot{\tilde{m}}+2r\ddot{\tilde{a}}\right)\right)+2a^2\left(\tilde{m}\left(\dot{\varphi }^2 \omega (\varphi )+2\varphi \ddot{\varphi }\right)+r \dot{\tilde{\varphi}} \dot{\varphi } \omega (\varphi )+2\varphi ^2\ddot{\tilde{m}}+\varphi \left(-2 \tilde{\varphi}'+3 \dot{\tilde{m}} \dot{\varphi }+r\ddot{\tilde{\varphi}}\right)\right)\right)\right) \label{gentpr}
\end{dmath}

\begin{dmath}
    \tilde{p}_t = \frac{1}{16 \pi  a^4 r \varphi ^2}\left(r\tilde{\varphi}\left(2 a^2 \dot{\varphi }^2 \omega (\varphi )+\varphi \left(2 \Gamma +2 a \dot{a} \dot{\varphi }+a^2\left(-\dot{\varphi }^2 \omega '(\varphi )+2\ddot{\varphi }\right)\right)\right)-\varphi \left(-a^4 r \varphi  \tilde{V}(\varphi )-4\varphi \left(\dot{a}^2 \tilde{m} \varphi +r\tilde{a}\left(\dot{a} \dot{\varphi }+\varphi \ddot{a}\right)\right)+2a\varphi \left(2\tilde{m}\left(\dot{a} \dot{\varphi }+2\varphi \ddot{a}\right)+r \left(\dot{a} \dot{\tilde{\varphi}}+2 \dot{\tilde{a}} \dot{\varphi }\right)-\varphi \left(\tilde{a}'+r \tilde{a}''-6 \dot{a} \dot{\tilde{m}}-2r\ddot{\tilde{a}}\right)\right)+2a^2\left(\tilde{m}\left(\dot{\varphi }^2 \omega (\varphi )+2\varphi \ddot{\varphi }\right)+r \dot{\tilde{\varphi}} \dot{\varphi } \omega (\varphi )+2\varphi ^2\ddot{\tilde{m}}-\varphi \left(\tilde{\varphi}'+r \tilde{\varphi}''-3 \dot{\tilde{m}} \dot{\varphi }-r\ddot{\tilde{\varphi}}\right)\right)\right)\right) \label{gentpt}
\end{dmath}

We are left with three unknown functions $\tilde{a}, \tilde{m}$ and $\tilde{\varphi}$, but only two equations \eqref{E(01)} and \eqref{phiFEepsilon}. To remedy this, we apply the additional constraint that the $\mathcal{O}(\epsilon)$ corrections to the tangential and radial pressures are equal, $\tilde{p}_t-\tilde{p}_r = 0$; that is, the matter is a perfect fluid at $\mathcal{O}(\epsilon)$, which yields

\begin{equation}
    a \left(r \tilde{\varphi}''-\tilde{\varphi}'\right)+\varphi  \left(r \tilde{a}''-\tilde{a}'\right) = 0
\end{equation}

\noindent This can be simply integrated to yield

\begin{equation}
    \tilde{a}(\eta,r) = \frac{R(\eta)}{2} r^2 + S(\eta) - \frac{a(\eta)}{\varphi(\eta)} \tilde{\varphi}(\eta,r) \label{perfluidsoln}
\end{equation}

\noindent The functions $R(\eta)$ and $S(\eta)$ are arbitrary functions coming from integration. To simplify the following analysis, we can set $R(\eta) = S(\eta) = 0$ by demanding the boundary condition that $\tilde{a} \to 0$ as $r \to \infty$ to obtain

\begin{equation}
    \tilde{a}(\eta,r) =  - \frac{a(\eta)}{\varphi(\eta)} \tilde{\varphi}(\eta,r) \label{perfluidsolnBC}
\end{equation}

\noindent Using this relation, we can define $p \equiv p_r = p_t$ to obtain

\begin{dmath}
    \tilde{\rho} = \frac{1}{16 \pi  a^4 r \varphi ^2}\left[r \tv \left(2 a^2 \dot{\varphi }^2 (\omega (\varphi )+3)+\varphi  \left(a \dot{\varphi } \left(6 \dot{a}-a \dot{\varphi } \omega '(\varphi )\right)+6 \Gamma \right)\right)+\varphi  \left(a^4 (-r) \varphi  \tilde{V}(\varphi )+2 a^2 \left(\varphi  \left(r \tv''-2 \varphi  \tm''+2 \tv'\right)-\dot{\varphi } \left(r \dot{\tv} (\omega (\varphi )+3)+\tm \dot{\varphi } \omega (\varphi )-3 \dot{\tm} \varphi \right)\right)+6 \dot{a} a \varphi  \left(-r \dot{\tv}+2 \dot{\tm} \varphi +2 \tm \dot{\varphi }\right)+12 \dot{a}^2 \tm \varphi ^2\right)\right]\label{perftrho}
\end{dmath}

\begin{dmath}
    \tilde{p} = \frac{1}{16 \pi  a^4 r \varphi ^2}\left[r\tv\left(2 a^2 \dot{\varphi }^2 (\omega (\varphi )+2)+\varphi \left(2 \Gamma -a\left(6 \dot{a} \dot{\varphi }+a\left(\dot{\varphi }^2 \omega '(\varphi )+2\ddot{\varphi }\right)\right)\right)\right)-\varphi \left(a^4 (-r) \varphi  \tilde{V}(\varphi )-4 \dot{a}^2 \tm \varphi ^2+2a\varphi \left(2\tm\left(\dot{a} \dot{\varphi }+2\varphi \ddot{a}\right)+3 \dot{a} \left(2 \dot{\tm} \varphi -r \dot{\tv}\right)\right)+2a^2\left(\tm\left(\dot{\varphi }^2 \omega (\varphi )+2\varphi \ddot{\varphi }\right)+r \dot{\tv} \dot{\varphi } (\omega (\varphi )+2)+2\varphi ^2\ddot{\tm}+\varphi \left(3 \dot{\tm} \dot{\varphi }-r\ddot{\tv}\right)\right)\right)\right] \label{perftp}
\end{dmath}

\noindent Then substituting in the values of $\tilde{\rho}$ and $\tilde{p}$ into \eqref{E(01)} and \eqref{phiFEepsilon} these two equations become

\begin{dgroup}[label={mainPDEs}]
\begin{dmath}
    -r \varphi  \left(\dot{a} r \tilde{\varphi}'+a r \dot{\tilde{\varphi}}'-a \dot{\varphi } \tilde{m}'\right)+2 \varphi ^2 \left(\dot{a} r \tilde{m}'+a r \dot{\tilde{m}}'-a \dot{\tilde{m}}\right)+a r^2 \dot{\varphi } \tilde{\varphi}' (\omega (\varphi )+2) \\ -\tilde{m} \varphi  \left(2 \dot{a} \varphi +a \dot{\varphi }\right) =0 \label{01eqn2}
\end{dmath}

\begin{dmath}
    a^3 r \varphi ^3 \tilde{V}'(\varphi )+2\varphi \left(-2 \dot{a} r \tilde{\varphi} \dot{\varphi } \left(\varphi  \omega '(\varphi )-3\right)+\varphi \left(\tilde{m}\left(-4 \dot{a} \dot{\varphi } \omega (\varphi )+6\varphi \ddot{a}\right) +\dot{a} \left(12 \dot{\tilde{m}} \varphi -r \dot{\tilde{\varphi}} (2 \omega (\varphi )+9)\right)\right)\right) +a\left(-r\tilde{\varphi}\left(\varphi 2 \varphi  \omega '(\varphi )-3\ddot{\varphi }+\dot{\varphi }^2 \left(\varphi ^2 \omega ''(\varphi )-\varphi  \omega '(\varphi )+8 \omega (\varphi )+12\right)\right)+2\varphi \left(\tilde{m}\left(\dot{\varphi }^2 \left(\omega (\varphi )-\varphi  \omega '(\varphi )\right)-2\varphi \omega (\varphi )\ddot{\varphi }\right)+2 r \dot{\tilde{\varphi}} \dot{\varphi } (2 \omega (\varphi )+3)-\varphi ^2\left(\tilde{m}''-3\ddot{\tilde{m}}\right)+\varphi \left(2 \tilde{\varphi}' (\omega (\varphi )+2)+r \tilde{\varphi}'' (\omega (\varphi )+2)-\dot{\varphi } \left(r \dot{\tilde{\varphi}} \omega '(\varphi )+4 \dot{\tilde{m}} \omega (\varphi )\right)-r\omega (\varphi )+3\ddot{\tilde{\varphi}}\right)\right)\right)=0 \label{phieqn2}
\end{dmath}
\end{dgroup}

\subsection{Second Perturbation}

We are only interested in a solution close to the bounce (that is to say, near $\eta=0$), which allows us to find a solution in the form of a truncated series expansion about $\eta=0$. Hence, we will attempt  to solve these differential equations, \eqref{01eqn2} and \eqref{phieqn2}, by writing $\tilde{m}$ and $\tilde{\varphi}$ as a series in time as follows:

\begin{dgroup}
\begin{dmath}
    \tilde{m}(\eta,r) = \tilde{m}_0(r)+\tilde{m}_1(r) \eta +\tilde{m}_2(r) \eta^2 + \tilde{m}_3(r) \eta^3 + \tilde{m}_4(r) \eta^4
\end{dmath}

\begin{dmath}    
    \tilde{\varphi}(\eta,r) = \tilde{\varphi}_0(r)+\tilde{\varphi}_1(r) \eta +\tilde{\varphi}_2(r) \eta^2 + \tilde{\varphi}_3(r) \eta^3 + \tilde{\varphi}_4(r) \eta^4
\end{dmath}
\end{dgroup}

\noindent Since \eqref{01eqn2} is a PDE that is first order in time and \eqref{phieqn2} is a PDE that is second order in time, this expansion will give us the general solution up to $\eta^4$ order with three arbitrary functions of $r$ chosen to be $\tm_0(r)$, $\tv_0(r)$ and $\tv_1(r)$. Substituting this series ansatz and solving for the functions of $r$ at each of the powers of $\eta$ gives the following solution:

\begin{dgroup}[label={genseriessoln}]

\begin{dmath}
    \tilde{m}_1(r) = B_0 r + B_1 \tilde{m}_0(r) + B_2 r \tilde{\varphi}_0(r) + B_3 r \tilde{\varphi}_1 (r)
\end{dmath}

\begin{dmath}
    \tilde{m}_2(r) = C_0 r +C_1 \tm_0(r) + C_2 r \tv_0(r) + C_3r \tv_1(r) + C_4 \tv_0'(r) + C_5 \tm_0''(r) \\ + C_6 r \tv_0''(r) + C_7 r \tilde{V}'(\varphi_0)
\end{dmath}

\begin{dmath}
    \tm_3(r) = D_0 r + D_1 \tm_0(r) + D_2 r \tv_0(r) + D_3 r \tv_1(r) + D_4 \tv_0'(r) + D_5 \tv_1'(r) + D_6 \tm_0''(r) + D_7 r \tv_0''(r) + D_8 r \tv_1''(r) + D_9 r \tilde{V}'(\varphi_0) + D_{10} r \tilde{V}''(\varphi_0)
\end{dmath}

\begin{dmath}
    \tm_4(r) = E_0 r + E_1 \tm_0(r) + E_2 r \tv_0(r)+ E_3 r \tv_1(r) + E_4 \tv_0'(r) + E_5 \tv_1'(r) + E_6 \tm_0''(r) + E_7 r \tv_0''(r) + E_8 r \tv_1''(r) + E_9 \tv_0^{(3)}(r) + E_{10} \tm_0^{(4)}(r) + E_{11} r \tv_0^{(4)}(r) + E_{12} r \tilde{V}'(\varphi_0) + E_{13} r \tilde{V}''(\varphi_0) + E_{14} r \tilde{V}^{(3)}(\varphi_0)
\end{dmath}

\begin{dmath}
    \tilde{\varphi}_2(r) = F_0 + \frac{F_1}{r} \tm_0(r) + F_2 \tv_0(r) + F_3 \tv_1(r) + \frac{F_4}{r} \tv_0'(r) + \frac{F_5}{r} \tm_0''(r) \\ + F_6 \tv_0''(r) + F_7 \tilde{V}'(\varphi_0)
\end{dmath}

\begin{dmath}
    \tilde{\varphi}_3(r) = G_0 + \frac{G_1}{r} \tm_0(r) + G_2 \tv_0(r) + G_3 \tv_1(r) + \frac{G_4}{r} \tv_0'(r) + \frac{G_5}{r} \tv_1'(r) + \frac{G_6}{r} \tm_0''(r) + G_7 \tv_0''(r) + G_8 \tv_1''(r) + G_9 \tilde{V}'(\varphi_0) + G_{10} \tilde{V}''(\varphi_0)
\end{dmath}

\begin{dmath}
    \tilde{\varphi}_4(r) = H_0 + \frac{H_1}{r} \tm_0(r) + H_2 \tv_0(r) + H_3 \tv_1(r) + \frac{H_4}{r} \tv_0'(r) + \frac{H_5}{r} \tv_1'(r) \\ + \frac{H_6}{r} \tm_0''(r) + H_7 \tv_0''(r) + H_8 \tv_1''(r) + \frac{H_9}{r} \tv_0^{(3)}(r) + \frac{H_{10}}{r} \tm_0^{(4)}(r) \\ + H_{11} \tv_0^{(4)}(r) + H_{12} \tilde{V}'(\varphi_0) + H_{13} \tilde{V}''(\varphi_0) + H_{14} \tilde{V}^{(3)}(\varphi_0)
\end{dmath}

\end{dgroup}

\noindent Here the capital coefficients $B_i, C_i,...$ are constants composed of the constants of the theory $\varphi_c$, $A$, $\Gamma$ and $\eta_0 $ as well as the constants from integration $c_1$, $c_2$, $c_3$ and $c_4$. The reason for presenting it as such is due to the length of this solution. One thing to note is that if all the integration constants $c_i$ are chosen to be zero, then all the capital constants with a zero subscript also vanish. Also, $\varphi_0 \equiv \varphi(0)$, which is given by 

\begin{equation}
    \varphi_0 = \varphi_c \left( 1 - \frac{2\eta_0 \Gamma}{A \lambda} \right)
\end{equation}

\noindent The conservation equations of matter, $\nabla_\mu T^{\mu \nu}=0$, at $\mathcal{O}(\epsilon)$ are given by

\begin{dgroup}
\begin{dmath}
    2 \pi  a^6 r \dot{\tilde{\rho}}+6 \pi  \dot{a} a^5 r \tilde{p} +6 \pi  \dot{a} a^5 r \tilde{\rho}+3 a^2 \Gamma  \dot{\tilde{m}}+3 a \Gamma  r \dot{\tilde{a}}-3 \dot{a} \Gamma  r \tilde{a} = 0
\end{dmath}

\begin{dmath}
    2 \pi  a^4 r^2 \tilde{p}'-\Gamma  r \tilde{m}'+\Gamma  \tilde{m} = 0
\end{dmath}
\end{dgroup}

\noindent Substituting the above solution and expanding up to $\eta^2$ order we find that the conservation equations are satisfied up to $\eta^2$ order.

\subsection{Boundary Conditions}

Since we are modelling an FLRW background with a central inhomogeneity, it is physically reasonable to apply the boundary condition that as we move farther from the central inhomogeneity the metric tends towards its FLRW form. That is to say, we want to assert the condition that as $r \to \infty$ we have $\tm(\eta,r) \sim \mathcal{O}(1)$ and $\tilde{a}(\eta,r) \sim \mathcal{O}(\frac{1}{r^\alpha})$ for $\alpha > 0$. This condition cannot uniquely define the arbitrary functions $\tm_0(r)$, $\tv_0(r)$ and $\tv_1(r)$; however, we can write these functions in the following general form that obeys the FLRW limit

\begin{equation}
    \tm_0(r) = \frac{d_0}{r^u} \sum_{n=1}^\infty \left(1+ \frac{x_n}{r^n} \right), \quad \tv_0(r) = \frac{d_1}{r^v} \sum_{n=1}^\infty \left(1+ \frac{y_n}{r^n} \right), \quad \tv_1(r) = \frac{d_2}{r^w} \sum_{n=1}^\infty \left(1+ \frac{z_n}{r^n} \right) \label{genArbFunc}
\end{equation}

\noindent for some coefficients $d_i$, $x_n$, $y_n$ and $z_n$ and constants $u$, $v$ and $w$. We can interpret our solution with the leading order behaviour in the limit $r \to \infty$, this allows us to take the leading order contributions from \eqref{genArbFunc} and make the following choice

\begin{equation}
    \tm_0(r) = \frac{d_0}{r^u} , \quad \tv_0(r) = \frac{d_1}{r^v}, \quad \tv_1(r) = \frac{d_2}{r^w} \label{bcansatz}
\end{equation}

\vspace{5mm}

\noindent which obey these boundary conditions for $u \geq0$ and $v,w \geq 1$ as well as requiring $\tilde{V}^{(n)}(\varphi_0)=0$ for $1 \leq n \leq 3$. Then the general perturbation to the potential up to $\eta^4$ order (which is the order up to which the solution \eqref{genseriessoln} is valid) is given by

\begin{dmath}
    \tilde{V}(\varphi) = V_0 \left( \varphi - \varphi_0 \right)^4 + V_1 \label{Vchoice}
\end{dmath}

\noindent This effectively removes the affect of the potential from the solution, except from the matter functions, since only derivatives of $\tilde{V}$ appear. If we take the limit $r \to \infty$ of \eqref{perftrho} and \eqref{perftp} combined with requiring $\tm(\eta,r) \to f(\eta)$ (as in solely a function of time) and $\tv (\eta,r) \to 0$ we find

\begin{dgroup}
\begin{dmath}
    \tilde{\rho} \to -\frac{\tilde{V}(\varphi)}{16 \pi}
\end{dmath}

\begin{dmath}
    \tilde{p} \to \frac{\tilde{V}(\varphi)}{16 \pi}
\end{dmath}
    
\end{dgroup}

\noindent We can interpret this result as an effective cosmological constant being introduced to the background cosmology due to the perturbation, since the perturbation to the energy density and pressure obey the equation of state $\tilde{p}=-\tilde{\rho}$ in the FLRW limit.

One way to constrain these functions is to demand that the solution is separable in the variables $\eta$ and $r$ like the solution found in \cite{YCL}. Immediately we require $v = w$, then substituting \eqref{bcansatz} into \eqref{genseriessoln} and requiring that all powers of $r$ be the same, we obtain two possibilities: i) $u=-1$ and $v=0$ ii) $u=0$ and $v=1$. Clearly, option i) does not obey the FLRW limit ($\tm = \mathcal{O}(1)$ as $r \to \infty$), hence, we proceed with the second option. In addition, ii) requires us to have the coefficients with a subscript `$0$' in \eqref{genseriessoln} to be zero, which is achieved by setting the integration constants to zero $c_1=c_2=c_3=c_4=0$.

We then obtain

\begin{equation}
    \tm_0(r) = d_0, \quad \tv_0(r) = \frac{d_1}{r}, \quad  \tv_1(r) = \frac{d_2}{r} \label{simplechoice}
\end{equation}

\noindent which implies that the functions for the perturbations must have the separable form

\begin{equation}
    \tm(\eta,r) = \beta(\eta), \quad \tv(\eta,r) = \frac{\gamma(\eta)}{r}, \quad \tilde{a}(\eta,r) =  - \frac{a(\eta)}{\varphi(\eta)} \frac{\gamma(\eta)}{r}
\end{equation}

\noindent We will consider the simple case \eqref{simplechoice} hereafter.

\subsection{Initial Conditions}

We have three constants coming from integration remaining $d_0$, $d_1$ and $d_2$. Two of these can be constrained by asserting some initial conditions on $\tv(\eta,r)$ (or $\tilde{a}(t,r)$). For example, requiring the perturbation to not effect the minimum of the bounce implies $\tilde{a}(0,r)=0$ and requiring the perturbation to maintain the parabolic nature of the bounce ($a \sim a_0 + a_2 t^2$) implies $\dot{\tilde{a}}(0,r)=0$. These conditions translate to $d_1 = d_2 =0$.

Applying these conditions to the simple case \eqref{simplechoice} we have

\begin{dgroup}[label={pertsolnIC}]
\begin{dmath}
    \tm_0(r) = d_0
\end{dmath}

\begin{dmath}
    \tm_1(r) = \frac{4 \Gamma ^2 d_0 \eta_0}{A^2-4 \Gamma ^2 \eta_0^2}
\end{dmath}

\begin{dmath}
    \tm_2(r) = \frac{2 \Gamma ^2 d_0 \left(A^4+28 A^2 \Gamma ^2 \eta_0^2-32 \Gamma ^4 \eta_0^4\right)}{\left(A^3-4 A \Gamma ^2 \eta_0^2\right)^2}
\end{dmath}

\begin{dmath}
    \tm_3(r) = \frac{32 \Gamma ^4 d_0 \eta_0 \left(7 A^4+16 A \Gamma  \eta_0 \lambda  \left(A^2-2 \Gamma ^2 \eta_0^2\right)+16 \Gamma ^4 \eta_0^4\right)}{3 A^2 \left(A^2-4 \Gamma ^2 \eta_0^2\right)^3}
\end{dmath}

\begin{dmath}
    \tm_4(r) = \frac{2 \Gamma ^4 d_0}{3 A^2 \left(A^2-4 \Gamma ^2 \eta_0^2\right)^4} \left(21 A^6+76 A^5 \Gamma  \eta_0 \lambda +16 A^4 \Gamma ^2 \eta_0^2 \left(35 \lambda ^2+97\right) -48 A^3 \Gamma ^3 \eta_0^3 \lambda -16 A^2 \Gamma ^4 \eta_0^4 \left(88 \lambda ^2+31\right)+1152 A \Gamma ^5 \eta_0^5 \lambda -1152 \Gamma ^6 \eta_0^6\right)
\end{dmath}

\begin{dmath}
    \tv_0(r) = 0
\end{dmath}

\begin{dmath}
    \tv_1(r) = 0
\end{dmath}

\begin{dmath}
    \tv_2(r) = \frac{1}{r} \frac{16 \Gamma ^4 d_0 \eta_0^2 \varphi_c \left(5 A^2-8 \Gamma ^2 \eta_0^2\right) (A \lambda -2 \Gamma  \eta_0)}{A^3 \lambda   \left(A^2-4 \Gamma ^2 \eta_0^2\right)^2}
\end{dmath}

\begin{dmath}
    \tv_3(r) = \frac{1}{r} \frac{64 \Gamma ^4 d_0 \eta_0 \varphi (A \lambda -2 \Gamma  \eta_0) \left(A^4+16 A \Gamma  \eta_0 \lambda  \left(A^2-2 \Gamma ^2 \eta_0^2\right)-30 A^2 \Gamma ^2 \eta_0^2+64 \Gamma ^4 \eta_0^4\right)}{3 \lambda   \left(A^3-4 A \Gamma ^2 \eta_0^2\right)^3}
\end{dmath}

\begin{dmath}
    \tv_4(r) = \frac{1}{r} \frac{16 \Gamma ^5 d_0 \eta_0 \varphi (A \lambda -2 \Gamma  \eta_0)}{3 A^3 \lambda  \left(A^2-4 \Gamma ^2 \eta_0^2\right)^4} \left(16 A^5 \lambda +A^4 \Gamma  \eta_0 \left(140 \lambda ^2+89\right)-740 A^3 \Gamma ^2 \eta_0^2 \lambda +4 A^2 \Gamma ^3 \eta_0^3 \left(131-88 \lambda ^2\right)+1696 A \Gamma ^4 \eta_0^4 \lambda -1568 \Gamma ^5 \eta_0^5\right)
\end{dmath}

\end{dgroup}

\noindent From this we can see that every term here has a factor of $d_0$ from which we can determine that $d_0$ can replace the artificial $\epsilon$ as the ``small" parameter under these initial conditions. A small $d_0$ would then imply that the mass of the central inhomogeneity is very small, $\tm \sim d_0$, which is consistent with the perturbative scheme where the central inhomogeneity is a small perturbation.

The solutions of the perturbation that is valid near $\eta = 0$ can be summarized by the equations

\begin{equation}
    \tm(\eta,r) = \beta(\eta), \quad \tv(\eta,r) = \frac{\gamma(\eta)}{r}, \quad \tilde{a}(\eta,r) =  - \frac{a(\eta)}{\varphi(\eta)} \frac{\gamma(\eta)}{r}
\end{equation}

\noindent with the functions $\beta(\eta)$ and $\gamma(\eta)$ given up to $\eta^4$ order by the coefficients of \eqref{pertsolnIC}. The matter functions are given by \eqref{perftrho} and \eqref{perftp} where $\tilde{V}$ is provided by \eqref{Vchoice}.

\subsection{Local Horizon}

We can compute the location of the horizon locally by solving the necessary condition $\theta_{(l)}\theta_{(n)}=0$ for $r=r_h(\eta)$,  where $\theta_{(l)}$ and $\theta_{(n)}$ are the outgoing/ingoing null expansion scalars. These expansions take the form

\begin{equation}
    \theta_{(\ell)}
    = \left(g^{\mu \nu}+ 2 l^{(\mu} n^{\nu)} \right) \nabla_{\mu}\ell_{\nu},
    \qquad
    \theta_{( n)}
    = \left(g^{\mu \nu}+ 2 l^{(\mu} n^{\nu)} \right) \nabla_{\mu}n_{\nu}
\end{equation}

\noindent where $l^\mu, n^\nu$ are the outgoing and ingoing null vectors, respectively, defined by

\begin{eqnarray}
    l^\mu &=& \left( \frac{1}{\sqrt{2}A_1(\eta,r)}, \frac{1}{\sqrt{2}A_2(\eta,r)},0,0 \right) \\ 
    n^\mu &=& \left( \frac{1}{\sqrt{2}A_1(\eta,r)}, -\frac{1}{\sqrt{2}A_2(\eta,r)},0,0 \right)
\end{eqnarray}

\noindent where $g_{00} = -(A_1)^2$ and $g_{11} = (A_2)^2$. Owing to spherical symmetry, the radial null congruences generated by $l^\mu$ and $n^\mu$ are shear-free. Moreover, they are hypersurface orthogonal and hence have vanishing rotation. Thus the only nontrivial optical scalar relevant to the present horizon calculation is the expansion.

Since we have a solution that is only valid locally, near the bounce, the best way to identify the black hole horizon is to find the future outer trapping horizon (FOTH) associated with the black hole whose definition is that on the outer horizon we have
\begin{eqnarray}
    \theta_{(l)} = 0 \\
    \theta_{(n)} < 0 \\
    \mathcal{L}_n \theta_{(l)} < 0
\end{eqnarray}

As is expected with black holes, we expect the radius of the horizon to be proportional to the mass. Since we have identified the parameter $d_0$ to be associated with the mass, and requiring the black hole horizon to be ``small", we use the ansatz 

\begin{dmath}
    r_h(\eta) = d_0 \Omega (\eta) + \Omega(d_0^{3/2})
\end{dmath}

\noindent and solve for $\Omega(\eta)$ using the condition $\theta_{(l)}=0$ then check the other two conditions. For the general McVittie metric \eqref{mcvittie}, we have

\begin{dmath}
    \theta_{(l)} = -\frac{\sqrt{2}}{A^2 a (M-2 r) (M+2 r)^3} \left[2 A^2 \dot{M} (M+2 r)^3+4 a r^2 A' \left(4 r^2-M^2\right)+A \left(\dot{A} (M+2 r)^4+4 a r (M-2 r) \left(M-2 r \left(M'+1\right)\right)\right)\right] \label{thetalgen}
\end{dmath}

\noindent Using the solution found with the simple case \eqref{pertsolnIC} and $\Omega(\eta) = \Omega_0 + \Omega_1 \eta + \Omega_2 \eta^2 + \mathcal{O}(\eta^3)$ we can expand \eqref{thetalgen} up to $\mathcal{O}(\eta^2)$ to obtain

\begin{dmath}
    \theta_{(l)} = \theta_{(l)}^{(0)} + \theta_{(l)}^{(1)} \eta + \theta_{(l)}^{(2)} \eta^2 + \mathcal{O}(\eta^3)
\end{dmath}

\noindent where $\theta_{(l)}^{(i)}$ are the coefficients of the Taylor expansion. We then solve the equations $\theta_{(l)}^{(i)} = 0$ up to leading order in $d_0$ for the coefficients $\Omega_i$. This is because we are only concerned with the value of $r_h$ up to $\mathcal{O}(d_0)$ to be in line with the rest of the analysis. We obtain

\begin{dgroup}
\begin{dmath}
    \Omega_0 = \frac{1}{2}
\end{dmath}

\begin{dmath}
    \Omega_1 = \frac{2 \Gamma ^2 \eta_0}{A^2-4 \Gamma ^2 \eta_0^2}
\end{dmath}

\begin{dmath}
    \Omega_2 = \frac{\Gamma ^2 \left(A^4-132 A^2 \Gamma ^2 \eta_0^2+224 \Gamma ^4 \eta_0^4\right)}{\left(A^3-4 A \Gamma ^2 \eta_0^2\right)^2}
\end{dmath}
    
\end{dgroup}

\noindent Therefore, we have the horizon radius given as a function of time up to $\mathcal{O}(\eta^2)$ and leading order in $d_0$

\begin{dmath}
    r_h(\eta) = \frac{d_0}{2} \left[ 1 +  \frac{4 \Gamma ^2 \eta_0}{A^2-4 \Gamma ^2 \eta_0^2} \eta + \frac{2\Gamma ^2 \left(A^4-52 A^2 \Gamma ^2 \eta_0^2+96 \Gamma ^4 \eta_0^4\right)}{\left(A^3-4 A \Gamma ^2 \eta_0^2\right)^2} \eta^2 + \mathcal{O}(\eta^3)  \right] + \mathcal{O}(d_0^{3/2}) \label{bhhorizon}
\end{dmath}

\noindent Note that the $d_0^{3/2}$ order terms should be included in the expression for $r_h$ to avoid a singularity in the null vectors at $\eta=0$, we choose not to present this for simplicity but have taken it into account. What is left to show is that the other two conditions for a FOTH are satisfied at both sides of the bounce, that is, the conditions are satisfied for $\eta \in [-\eta_*,\eta_*]$ for some $\eta_* > 0$. At the bounce, $\eta = 0$, at leading order in $d_0$ (which is the largest contribution to respective functions) we obtain

\begin{dmath}
    \theta_{(n)} \big|_{\eta = 0} = -\frac{8 \Gamma ^{3/2}\sqrt{ \eta_0 \varphi_c (A \lambda -2 \Gamma \eta_0) }}{\sqrt{A \lambda d_0} |4 \Gamma ^2 \eta_0^2 - A^2| }
\end{dmath}

\begin{dmath}
    \mathcal{L}_n \theta_{(l)} \big|_{\eta = 0} = - \frac{\Gamma  \varphi_c (A \lambda -2 \Gamma  \eta_0)}{A d_0^2 \lambda  \left(4 \Gamma ^2 \eta_0^2 - A^2 \right)}
\end{dmath}

\noindent Both are negative for $\frac{2 \Gamma \eta_0}{\lambda} < A < 2 \Gamma \eta_0$ and $A,\varphi_c, \eta_0 , d_0 > 0$ which is consistent with the valid conditions in section \ref{sec2.1}. Hence, since these functions are negative at the minimum of the bounce, $\eta=0$, then there exists some $\eta_* > 0$ such that $\theta_{(n)} < 0$ and $\mathcal{L}_n \theta_{(l)} < 0$ for $\eta \in [-\eta_*,\eta_*]$. This implies the horizon given by \eqref{bhhorizon} is indeed a future trapping horizon on both sides of the bounce, near the bounce in time.

In addition, we can see that when $\Gamma \to 0$ (which forces $A \to 0$ to satisfy the FOTH condition $\frac{2 \Gamma \eta_0}{\lambda} < A < 2 \Gamma \eta_0$), the matter terms of all orders vanish and, in fact, looking at \eqref{pertsolnIC}, all the $\epsilon^1$ order terms vanish except for the ones of order $\mathcal{O}(\eta^0)$. Consequently, the metric reduces to the Schwarzschild metric in isotropic coordinates. In this same limit, the horizon \eqref{bhhorizon} becomes the expected Schwarzschild horizon $r_h = \frac{d_0}{2}$.

It is important to note that we cannot be certain that the horizon \eqref{bhhorizon} is a black hole horizon due to the limitations of the perturbative scheme. This analysis is only valid when $d_0 \ll 1$ and $\eta \ll 1$ while an event horizon requires global knowledge of the spacetime. However, it is clear that the small horizon is associated with the central inhomogeneity as it vanishes when $ d_0 \to 0$ which causes the mass $\tm \sim d_0$ to vanish.

\section{Discussion}

We have found a perturbative solution in scalar-tensor theory which models a black hole in an evolving cosmological background, with the goal of studying black hole persistence. A modified gravity theory, like scalar-tenor theory, provides us with a mathematical and physical justification to model a bounce. However, once we attempt to model a central inhomogeneity embedded within a cosmology, the equations become increasingly complicated, especially within a modified theory of gravity. Therefore, we use a perturbative scheme where the leading order field equations are solved by a bouncing FLRW model with flat spatial curvature and radiation perfect fluid \cite{barrow_nonsingular_1993}. The second order terms introduce a central inhomogeneity using the generalized McVittie metric \eqref{mcvittie}. In the perturbative scheme \eqref{pertansatzA}-\eqref{pertansatzPhi}, written with perturbative parameter $\epsilon$, the $\mathcal{O}(\epsilon^0)$ field equations are satisfied by the FLRW cosmological solution, while the $\mathcal{O}(\epsilon)$ field equations are solved for the functions representing the perturbations (expressed by the functions with a tilde).We start with an imperfect fluid with radial and tangential pressures which, combined with the energy density, solves the diagonal diagonal field equations. We then assert that the perturbations represent a perfect fluid, which allows us to solve for $\tilde{a}$ by relating it to $\tilde{\varphi}$. We are then left to solve for two field equations, \eqref{01eqn2} and \eqref{phieqn2}, for the two functions $\tm$ and $\tv$, which we do as a series solution up to $\mathcal{O}(\eta^4)$ which is valid near the bounce at $\eta =0$. The general series solution introduces three arbitrary functions of $r$ ($\tm_0(r)$, $\tv_0(r)$ and $\tv_1(r)$), which are constrained by demanding the McVittie metric \eqref{mcvittie} asymptotes to the FLRW metric in the limit $r \to \infty$. We then choose certain initial conditions that maintain the parabolic nature of the bounce, from which we find that the constant of integration, $d_0$, is the true small perturbative parameter as when $d_0 \to 0$ the perturbations all vanish. 

We have found a small ($r_h \sim d_0$) evolving horizon which we have interpreted as the horizon of the of the central inhomogeneity as when the mass term vanishes, $d_0 \to 0$, the horizon vanishes and in the limit that the matter becomes a vacuum $\Gamma \to 0$, $d_0$ plays the role of the Schwarzschild mass. We also show that this horizon is a future outer trapping horizon near the bounce $\eta \sim 0$. The existence of this horizon near the minimum of the bounce characterizes the persistent nature of the black hole. We note that the evolution of the horizon due to the cosmological coupling is not symmetric about the bounce (due to the linear term in $r_h$ in terms of $\eta$.).

\section*{Declarations}

\subsection*{Funding}

B.Y. acknowledges support from the Department of Mathematics and Statistics at Dalhousie University, Canada. A.A.C. is supported by the Natural Sciences and Engineering Research Council of Canada (NSERC).  

\subsection*{Financial Interests}

The authors have no relevant financial or non-financial interests to disclose.

\subsection*{Conflict of Interest}

A.A.C. is an Associate Editor of the journal General Relativity and Gravitation, but was not involved in the peer review of this article.

\end{document}